\documentstyle[12pt,aaspp4]{article}

\def\mc{\markcite}
\def\hd{HD 98800}
\def\cd{CD $-$33$\arcdeg$7795}
\def\cdb{CD $-$29$\arcdeg$8887}
\def\hen{Hen 3--600}
\def\rosat{{\it ROSAT\/}}
\def\iras{{\it IRAS\/}}
\def\hri{{\rm HRI}}
\def\etal{et al.}
\def\rne{Neuh\"auser}
\def\Lx{\mbox{$L_{\rm X}$}}
\def\Lbol{\mbox{$L_{\rm Bol}$}}
\def\lxlbol{\Lx/\Lbol}
\def\sec{Sec.}
\def\pms{pre--main-sequence}
\def\per{\mbox{$^{-1}$}}
\begin{document}

\title{ROSAT and Hipparcos Observations of Isolated Pre--Main-Sequence
  Stars near \hd}

\author{Eric L. N. Jensen,\altaffilmark{1} David H. Cohen,\altaffilmark{2}
\& Ralph Neuh\"auser\altaffilmark{3}
}

\altaffiltext{1}{Arizona State University, Department of Physics \&
  Astronomy, P.O. Box 871504, Tempe, AZ 85287-1504 USA.
  Email:  Eric.L.N.Jensen@asu.edu}
\altaffiltext{2}{University of Wisconsin
-- Madison,  Department of Astronomy, 475 N.\ Charter Street, and Fusion 
Technology Institute, 1500 Engineering Drive, Madison,
 WI 53706 USA. Email:  cohen@duff.astro.wisc.edu}
\altaffiltext{3}{Max-Planck-Institut f\"ur extraterrestrische Physik,
  Giessenbachstra{\ss}e 1, D-85740 Garching, Germany.
  Email: rne@rosat.mpe-garching.mpg.de}

\begin{abstract}
  
  We present new observations of the isolated young stars \hd\ and
  \cd.  Pointed \rosat\ observations show that their X-ray properties,
  including X-ray luminosity and variability, are consistent with
  those of \pms\ (PMS) stars.  These observations do not reveal any
  additional PMS candidates in 40\arcmin\ fields centered on \hd\ and \cd.
  Hipparcos observations of TW Hya (Wichmann \etal\ 1998\mc{w98}) and
  \hd\ (Soderblom \etal\ 1998\mc{s98}) show that both stars are
  roughly 50 pc away and are PMS with ages of $\sim 10^7$ yr.  We
  searched the Hipparcos catalog (complete down to $\sim$ 2--3 $L_\odot$ at
  this distance) for other PMS stars in the same area.
  In a 10-pc radius volume of space centered on the previously known
  PMS stars, we find one additional candidate PMS star (CD
  $-$36$\arcdeg$7429) with a low space velocity, X-ray emission
  comparable to that of \hd, and Li absorption.  There are eight other
  stars in this area that have dwarf spectral types and lie above the
  main sequence, but based on their weak X-ray emission, high space
  velocities, and lack of Li in low-resolution spectra (i.e.\ EW(Li)
  $< 0.1$ \AA), these are probably mis-classified subgiants or giants.
  The current positions and proper motions of TW Hya, \hd, and CD
  $-$36$\arcdeg$7429 are inconsistent with them having formed as a
  group.
  
\end{abstract} 

 
\section{Introduction} 

Recent observations have revealed a small population
of stars that bear many of the hallmarks of low-mass
pre--main-sequence stars but lie far from any obvious region of recent
star formation (as revealed by substantial dark clouds of molecular
gas and dust).  Gregorio-Hetem \etal\ (1992\mc{gh92}) identified 33
candidate T Tauri stars based on
spectroscopy of stars in the {\it IRAS\/} Point Source Catalog and a few additional emission-line stars.  One of the more interesting
findings of their work was the identification of four stars (HD 98800,
\cd, \cdb, and \hen) within 10\arcdeg\ of TW Hya, earlier
identified as a possible isolated T Tauri star by Rucinski \& Krautter
(1983\mc{rk83}).  The proximity of these five systems to each other
suggested a possible loose cluster of young stars, possibly formed by
a small molecular cloud that has since dissipated (e.g., Feigelson
1996\mc{f96}).

In the absence of association with any known cloud complex, the
distances to these stars were very uncertain, and thus their
pre--main-sequence status was in question. To investigate whether two
of these stars, \hd\ and \cd, are in fact
young and whether they are part of a larger group of young stars, we
observed them with the \rosat\ X-ray satellite.  One distinguishing
feature of low-mass pre--main-sequence stars is strong X-ray emission,
presumed to arise from solar-like chromospheric activity (see, e.g.,
\rne\ 1997\mc{n97} for a recent review).  Thus, X-ray observations can
be used to search for young stars that may not have other hallmarks of
youth such as infrared excesses or strong H$\alpha$ emission.
Observations of Taurus-Auriga with the {\it Einstein} satellite led to
the discovery of the naked T Tauri stars, stars that have no
circumstellar material but that are nonetheless coeval with classical
T Tauri stars (Walter 1986\mc{w86}).

Much observational attention has been focused on the stars in the
vicinity of TW Hya recently, and several other investigations have
proceeded in parallel with ours.  Kastner \etal\ (1997\mc{k97})
reported \rosat\ observations of some of the same stars we report on
here.  They found that the strength of X-ray emission from these stars
is consistent with them being young stars and argued that the five
young stars in this area make up a physical association.  Hoff \etal\ 
(1996\mc{hph96}, 1998\mc{hhp98}) investigated TW Hya and \cdb\ and
their surroundings with low spatial resolution using \rosat\ PSPC
pointed observations.  Soderblom \etal\ (1998\mc{s98}) and Wichmann
\etal\ (1998\mc{w98}) report Hipparcos distances to HD 98800 and TW
Hya, respectively; both systems have ages of $\sim$ 1--2 $\times\ 
10^7$ yr and distances of $\sim 50$ pc.  The work we report here is
complementary to this other recent work. We analyze the X-ray data of
our target stars in more detail than Kastner \etal\ (1997\mc{k97}) and
explore the status of other X-ray sources in the surrounding fields.
We also combine X-ray and Hipparcos data to search for other young
stars in the vicinity in order to address the larger question of the
origin of these isolated young stars.

In \sec\ \ref{sec:observations} we report the details of our
observations. In \sec\ \ref{sec:xproperties} we show that the X-ray
properties of the two target stars are consistent with their being
\pms\ (PMS) stars.  We then investigate (\sec\ \ref{sec:othermembers})
whether there is any evidence that the young stars near TW Hya formed
as a group, using our X-ray observations as well as data from the
Hipparcos satellite.  Finally, in \sec\ \ref{sec:discussion} we
discuss the implications of our findings for our understanding of
isolated young stars.

\section{\rosat\/ HRI Observations}\label{sec:observations}

Observations of \hd\ and \cd\ were made with the High Resolution
Imager (\hri) aboard \rosat.  The telescope consists of four nested
Wolter type I mirrors with a peak effective area of almost 1000 cm$^2$
(Tr\"umper 1983\mc{t83}).  The \hri, described in David \etal\ 
(1997\mc{d97}), uses a two microchannel plate detector with a CsI
photocathode.  The mirror and detector combination is sensitive to
X-rays with energies between 0.1 keV and 2.4 keV.  Its very good
spatial resolution is characterized by a response function having an
on-axis FWHM of about 5\arcsec.  It has a field of view with an
approximate diameter of 40\arcmin.

With a low and well understood background, \rosat\/ has a sensitivity
that is the best of any current or past large X-ray telescope.  The
minimum detectable flux in our observations is about $10^{-14}$ ergs
s$^{-1}$ cm$^{-2}$.  The sensitivity declines slightly with off-axis
angle, reaching 85 percent of the on-axis value at 20\arcmin\/ for 1
keV photons.  Because of the good sensitivity and spatial resolution
the \rosat\/ \hri\/ is well suited to identifying point sources in
potentially crowded fields.

The pointed \hri\/ observations were made consecutively over a period
of 20 days.  The \hd\ observations took place from 1994 December 28 to
1995 January 7 with an effective exposure time of 23,051 s.  The \cd\ 
observations took place from 1995 January 9 to 17 with an effective
exposure time of 30,604 s.

In both fields the central stars (\hd\ and \cd, respectively) are the
brightest X-ray sources, but in each case several other sources are
also visible.  To locate all statistically significant sources we used
the source detection algorithms in the MIDAS/EXSAS software package
(January 1995 version, Zimmermann \etal\ 1993\mc{z93}).  These
algorithms use a three-step process in which sources first are located
using a sliding window technique.  Next a background map is made by
excluding these sources and the sliding window search is repeated
using the new background determination.  Finally a maximum likelihood
statistic ${\cal L}$ is calculated for each possible point source
using the background map and knowledge of the point spread function.
We used a maximum likelihood threshold value of 8, which corresponds
to a probability that a spurious source is identified at a given
position of $3.3 \times 10^{-4}$.

Before extracting the identified sources we examined the radial
profiles.  In general the source count rate became indistinguishable
from the background level by 20\arcsec.  We chose a radius for each
source extraction circle accordingly.  We then extracted a background
sample from a nearby, source-free region of the detector.  The
background count rate was scaled to the area of the source extraction
circle and was subtracted from the source.  The net count rate was
then computed by summing over all 16 \hri\/ energy channels.

In the \hd\ field we detected nine X-ray sources, and in the \cd\ 
field we detected six sources.  In Tables \ref{table:sources-hd} and
\ref{table:sources-cd} we list the sources detected in the two fields,
their positions, count rates, and fluxes.  In each field the central
star is more than an order of magnitude brighter in X-rays than any of
the other sources; there are no X-ray sources in these fields of
comparable brightness to the previously-known PMS stars HD 98800 and
\cd.

\begin{deluxetable}{ccc}
\tablecaption{X-ray sources in the HD 98800 Field \label{table:sources-hd}}
  \tablewidth{0pt}
  \tablenum{1}
\tablehead{
 \colhead{Source} & \colhead{Count
  Rate} & \colhead{X-ray Flux} \\
  \colhead{} & \colhead{(counts ks$^{-1}$)} &
  \colhead{($10^{-14}$ erg s$^{-1}$ cm$^{-2}$)}
}
\startdata
RX J112205.4$-$244641\tablenotemark{a} & $127\phd\phn\phn \pm 2\phd\phn\phn$ &  $326 \pm 49$\\ 
 RX J112303.2$-$243629  &  
  $\phn\phn1.52  \pm 0.44$ & $4.0 \pm 1.3$\\
 RX J112157.4$-$244014   &
  $\phn\phn0.41 \pm 0.18$ & $1.1 \pm 0.5$\\
 RX J112112.4$-$244032    &
  $\phn\phn1.23 \pm 0.38$ & $3.2 \pm 1.1$\\
 RX J112229.2$-$244258    &
  $\phn\phn1.66 \pm 0.33 $ & $4.3 \pm 1.1$\\
 RX J112154.5$-$244413    &
  $\phn\phn0.94 \pm 0.54$ & $2.4 \pm 1.4$\\
 RX J112204.4$-$245347   &
  $\phn\phn3.79 \pm 0.43$ & $9.7 \pm 1.8$\\
 RX J112156.1$-$245503   &
  $\phn\phn1.52 \pm 0.32$ & $3.9 \pm 1.0$\\
 RX J112216.6$-$245810   &
  $\phn\phn0.82 \pm 0.26$ & $2.1 \pm 0.7$\\
\enddata
\tablenotetext{a}{HD 98800}
\end{deluxetable}

\begin{deluxetable}{ccc}
  \tablewidth{0pt}
  \tablenum{2}
  \tablecaption{X-ray sources in the CD $-$33$\arcdeg$7795
    Field \label{table:sources-cd}}
  \tablehead{
    \colhead{Source} & \colhead{Count Rate} & \colhead{X-ray Flux} \\
    \colhead{} & \colhead{(counts ks$^{-1}$)} &
    \colhead{($10^{-14}$ erg s$^{-1}$ cm$^{-2}$)}
    }
  \startdata
  RX J113155.3$-$343630\tablenotemark{a}
  & $156\phd\phn\phn \pm 2\phd\phn\phn$ & $400 \pm 60\phd\phn$ \\
  RX J113220.7$-$343343
  & $\phn\phn0.60 \pm 0.22$ & $1.6 \pm 0.6$ \\ 
  RX J113126.2$-$343411
  & $\phn\phn0.81 \pm 0.18$ &  $2.1 \pm 0.6 $ \\ 
  RX J113143.3$-$343704
  & $\phn\phn0.46 \pm 0.26$ &  $1.2 \pm 0.7 $ \\ 
  RX J113213.6$-$344127
  & $\phn\phn1.52 \pm 0.28$ &  $3.9 \pm 0.9 $ \\ 
  RX J113140.5$-$344246
  & $\phn\phn0.50 \pm 0.21$ &  $1.3 \pm 0.6 $ \\ 
  \enddata
  \tablenotetext{a}{CD $-$33$\arcdeg$7795}
\end{deluxetable}

The conversion of X-ray count rates to fluxes is model dependent
because the effective area of the \rosat\/ \hri\/ is a strongly
varying function of photon energy.  Late-type stars have optically
thin thermal X-ray spectra that are typically fit by isothermal ($T
\approx 10^7$ K) models (Schmitt \& Snowden 1990\mc{ss90}).  The
reddening of both \hd\ and \cd\ is negligible, implying that
interstellar attenuation of the X-rays is also negligible.  We derived
the fluxes of the central stars of the two fields assuming a Raymond
and Smith isothermal model having a temperature of $10^7$ K and
attenuated by an interstellar column of $N_H=5 \times 10^{19}$
cm$^{-2}$.  This procedure effectively gave a count rate to flux
conversion factor, which we then used for each of the other stars in
the fields.  This factor, of roughly $2.5 \times 10^{-11}$ ergs s\per\ 
cm$^{-2}$ count\per, is in good agreement with the values listed in
the \hri\ handbook (David \etal\ 1997).  We note that the count rate
to flux conversion factor varies by no more than 30 percent over the
range of spectral parameters $0.5 \lesssim T_X \lesssim 1.0$ keV and
$10^{19} \lesssim N_H \lesssim 10^{20}$ cm$^{-2}$.  This range of
$N_H$ values corresponds to visual extinctions of $A_V \approx 0.005$
to 0.05 magnitudes, consistent with the observed low extinction toward
the central sources in our fields.

We use this same value to convert the fluxes in all of the sources.
This assumption will be valid if the other sources are PMS stars or
late-type main sequence stars at a distance similar to the central
star in each field.  If the sources are background sources then their
intrinsic fluxes will tend to be higher than our estimates, due to the
probable underestimate of the interstellar absorption.  The fluxes are
listed in Tables \ref{table:sources-hd} and \ref{table:sources-cd};
the flux uncertainties quoted are derived from the uncertainties in
the count rates added in quadrature with an assumed 15\% uncertainty
in the count rate to flux conversion.

\section{X-ray Properties of \hd\ and \cd}\label{sec:xproperties}

\subsection{X-ray Fluxes}\label{sec:xflux}

The X-ray fluxes of \hd\ and \cd\ combined with their $V$ magnitudes
(Soderblom \etal\ 1998\mc{s98}, Gregorio-Hetem \etal\ 1992\mc{gh92})
and bolometric corrections (Kenyon \& Hartmann 1995\mc{kh95}) give
\lxlbol\ ratios of $2.8 \times 10^{-4}$ and $1.2 \times 10^{-3}$,
respectively.\footnote{The value of \lxlbol\ for \hd\ is an upper
  limit for \hd\ Aa (the brightest star in this quadruple system,
  Torres \etal\ 1995\mc{t95}) that results from assigning all of the
  X-ray flux to Aa and using the $V$ magnitude of this star from
  Soderblom \etal\ (1998\mc{s98}). The true \lxlbol\ ratios of the
  individual stars in the system are highly uncertain because it is
  unknown how the X-ray flux is divided among the four stars.}  The
X-ray luminosity of \hd\ given its distance of 46.7 pc (Soderblom
\etal\ 1998\mc{s98}) is $8.5 \times 10^{29}$ erg s\per\ cm$^{-2}$, or
$\log \Lx = 29.9$.  Assuming the same distance for \cd\ gives $\log
\Lx = 30.0$.  The strength of X-ray emission from \hd\ and \cd\ is
consistent with the interpretation that they are PMS stars; their
X-ray luminosities are within the ranges found by \rne\ \etal\ 
(1995\mc{n95}) for T Tauri stars in Taurus and are higher than all but
the few most X-ray-luminous stars in the Pleiades or Hyades (\rne\ 
1997\mc{n97}).  Even if the X-ray luminosity of \hd\ is divided
equally among the four stars in the quadruple system (Torres \etal\ 
1995\mc{t95}), the X-ray luminosity of each star is comparable to that
from PMS stars.  The \lxlbol\ values are also in the range found for
Taurus PMS stars (\rne\ \etal\ 1995\mc{n95}).

\subsection{Time Variability}\label{sec:variability}

The observations of \hd\ and \cd\ each span about ten days, and
because of their relatively high count rates we can address the issue
of time variability using these data.  Because \rosat\/ is in
low-Earth orbit the data are in sections of $t \approx 2500$ s,
separated by slightly longer (and occasionally much longer) intervals.
Individual orbits provide convenient time bins for tests of
variability on timescales of hours.  We also divided the data into 100
s bins to test for shorter-term variability.

\begin{figure}
  \plottwo{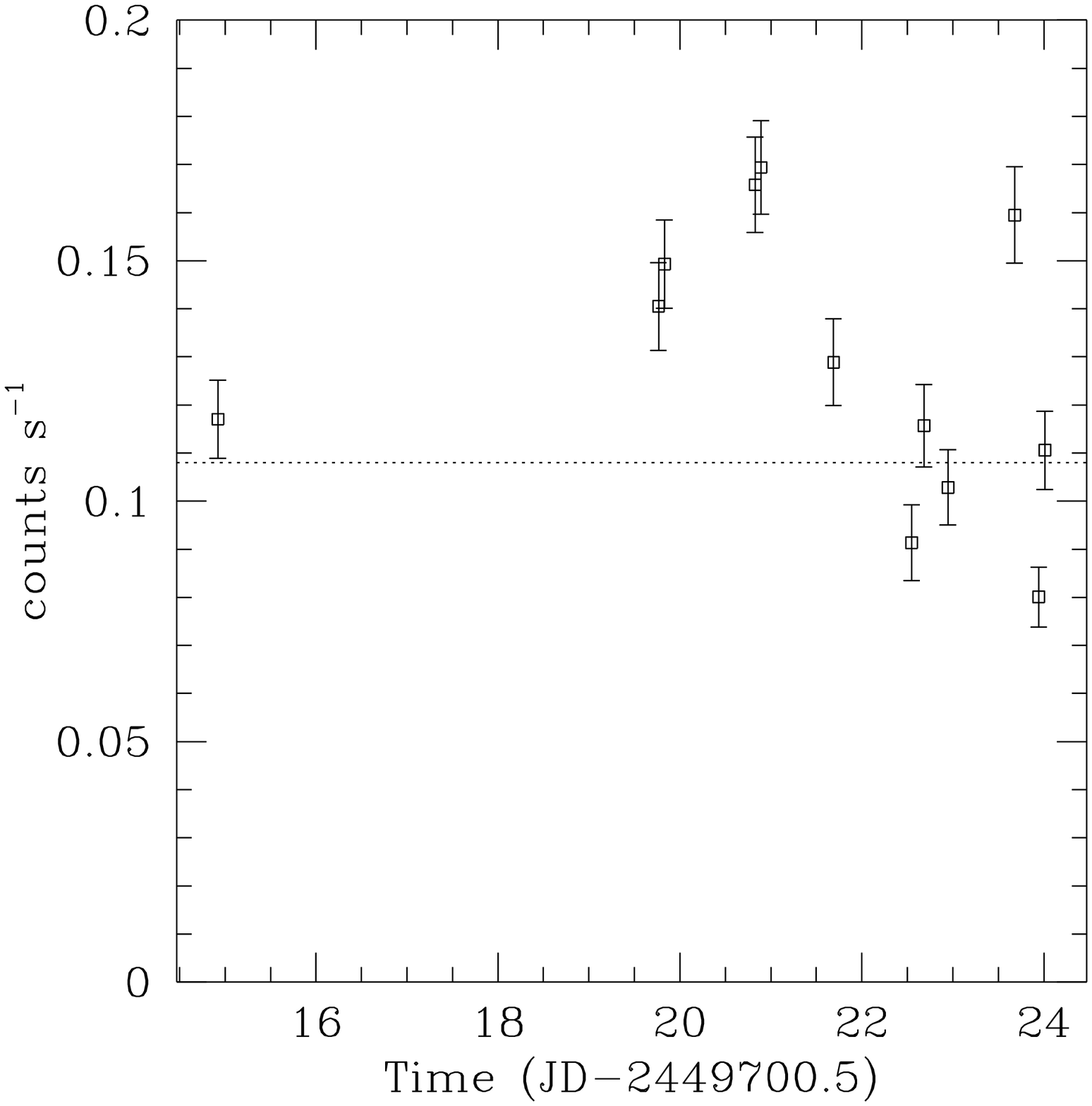}{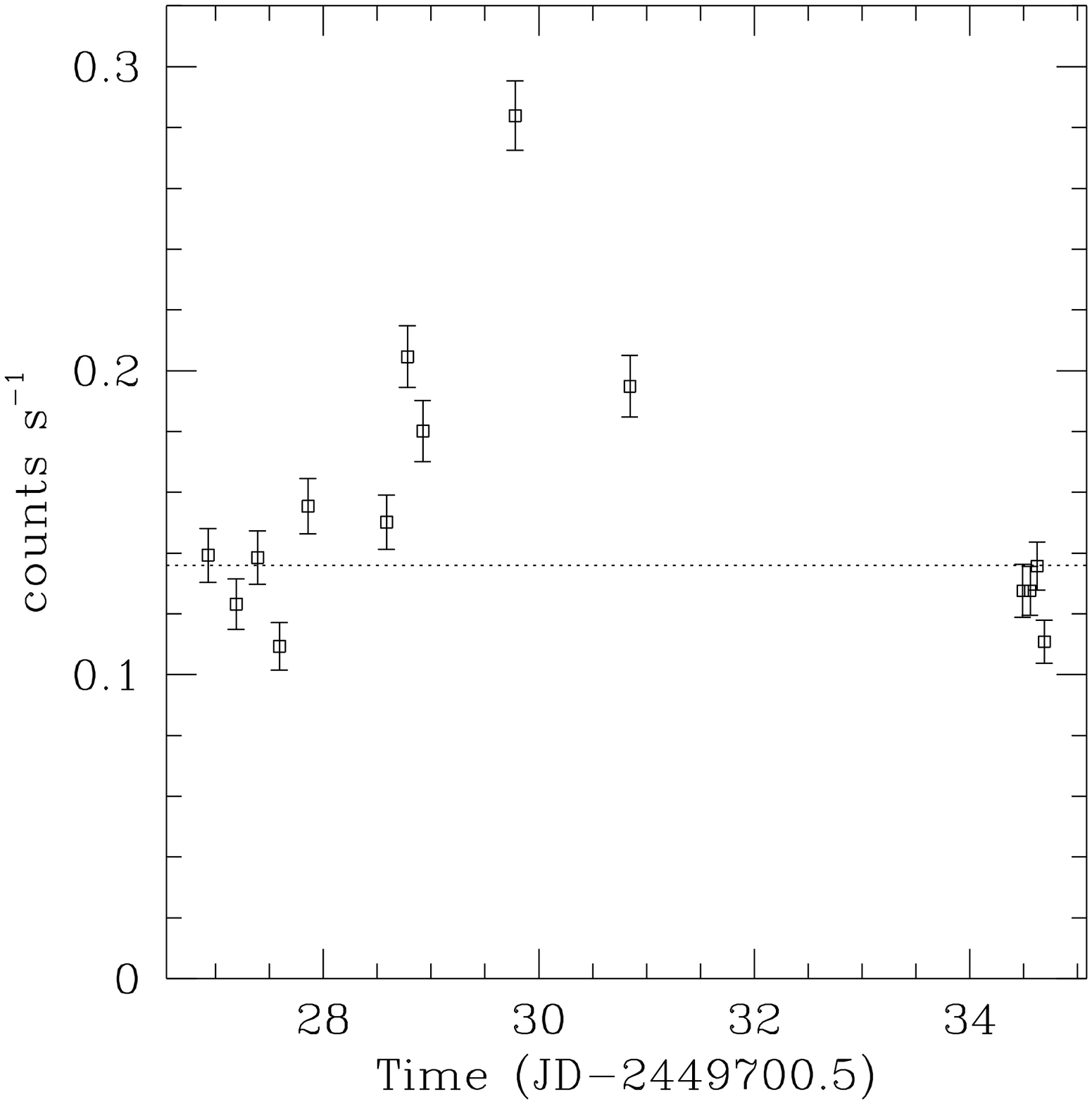}
\caption{X-ray light curves of
  \hd\ (left) and \cd\ (right).  Each point represents one orbit, or
  approximately 2500 s of effective observing time.  The dashed lines
  indicate the mean count rates for each target excluding the flare
  events.
  \label{figure:lc}
  }
\end{figure}

In Figure \ref{figure:lc} we show the light curves for each of the two
stars with the 2500 s bins.  In both stars there is a flare-type event
with a rise time of approximately one day.  For \cd\ the amplitude of
the event is at least 90 percent above the mean count rate.  For \hd\ it
is at least 40 percent.  In both cases the \rosat\ count rates returned
to the pre-flare levels within one day of reaching their peaks.

There is very little variability on shorter timescales as determined
from the light curves with the 100 s bins.  We fit the light curves
from each separate orbit with a constant source model.  In almost
every case the data from each orbit are consistent with a constant
source flux, although this constant source flux level is different for
each orbit.  For \hd\ only 2 out of 12 orbits show short term
variability with a significance greater than 95 percent.  For \cd\ 
only 1 out of 14 orbits shows short term variability above this
significance level.  The standard deviation of the source counts in
the 100 s bins for each orbit is typically 20 percent of the mean
count rate for that orbit, which is similar to the statistical
uncertainties.  We can therefore conclude that very little variability
with amplitudes greater than 20 percent is seen on timescales of
minutes, and none of this short timescale variability has an amplitude
greater than 40 percent.

While X-ray flares are seen in late-type stars of all ages (Schmitt
1994\mc{s94}), the presence of flares in both of our targets within a
relatively short time suggests that these stars are part of an active
population.  Gagn\'e \etal\ (1995\mc{gcs95}) estimated that late-type
Pleiades cluster members (ages $\sim 7 \times 10^7$ yr) undergo large
flares roughly 1\% of the time.  Thus, if our target stars have a
similar flaring frequency, the probability that we would observe a
$\sim$ 1 day flare in a single 10-day observation of a given star was
$\sim$ 10\%.  The probability that we would observe flares in {\em
  both\/} stars was about 1\%.  Thus, the observations suggest a
somewhat higher flaring frequency than that of the Pleiades,
consistent with younger ages for these stars.  \rne\ \& Preibisch
(1995\mc{np95}) estimate an X-ray flare rate of roughly one flare
every 40 days for T Tauri stars in Taurus-Auriga.  If the flare rate
for \cd\ and \hd\ is similar to that of T Tauri stars, this gives a
$\sim$ 25\% probability that we would see a flare in a 10-day
observation of one star, and a $\sim$ 6\% probability of seeing flares
in both. This indicates that \hd\ and \cd\ are at least as active as T
Tauri stars, though the X-ray flare rate determined by \rne\ \&
Preibisch (1995\mc{np95}) is very uncertain.  We conclude that the
level of X-ray activity in \hd\ and \cd\ is consistent with their
being young stars.

\subsection{Spectral Properties}\label{sec:spectral}

Although the \rosat\ \hri\/ data sets do not afford us the opportunity
to perform quantitative X-ray spectral fitting, we were able to
perform some spectral modeling using the shorter PSPC observations of
\hd\ and \cd\ taken from the \rosat\/ All-Sky Survey (RASS).  With the
RASS, the whole sky has been observed with the PSPC (Positional
Sensitive Proportional Counter), which offers moderate spectral
resolution.  The RASS fluxes of the previously known five PMS stars in
this field have been presented by Kastner \etal\ (1997\mc{k97}).
However, they do not list hardness ratios, which are basic spectral
information derived from the relative fluxes in the different energy
bands of the PSPC (\rne\ \etal\ 1995\mc{n95})\footnote{
  If $Z_{\rm s}$, $Z_{\rm m}$, and $Z_{\rm h}$ denote count rates in the
soft (0.1 to 0.4 keV), medium (0.5 to 0.9 keV),
and hard (0.9 to 2.1 keV) PSPC bands respectively, then
${\rm HR}~1~=~\frac{ Z_{\rm h} + Z_{\rm m} - Z_{\rm s} } { Z_{\rm h} + Z_{\rm m} + Z_{\rm s} }$
and
${\rm HR}~2~=~\frac{ Z_{\rm h} - Z_{\rm m} } { Z_{\rm h} + Z_{\rm m} }$;
both ratios can range from $-1$ to $+1$.
}.  Therefore, we include
the basic spectral analysis below for all five stars.

In Table \ref{table:rasspms}, we list the five previously known PMS
stars with their maximum likelihood ${\cal L}$ of existence as X-ray
sources in the RASS, as well as the RASS exposure times, hardness
ratios (HR 1 and HR 2), and X-ray fluxes, derived using an energy
conversion factor of 1 count = $(5.30 \times {\rm HR1} + 8.31) \times
10^{-12}$ erg s\per\ cm$^{-2}$ (Schmitt \etal\ 1995\mc{sfg95}).  The
hardness ratios were derived using the procedure given in Neuh\"auser
\etal\ (1995\mc{n95}).

\begin{deluxetable}{lccccc}
  \tablewidth{0pt}
  \tablenum{3}
  \tablecaption{X-ray spectral properties of previously-known
    young stars near TW Hya
    \label{table:rasspms}
    }
  \tablehead{
    \colhead{Name} & \colhead{${\cal L}$} & \colhead{Exposure} &
    \colhead{HR 1} & \colhead{HR 2} & \colhead{Flux/$10^{-12}$ } \\
    \colhead{} & \colhead{} & \colhead{Time (s)} & \colhead{} &
    \colhead{} & \colhead{erg s$^{-1}$ cm$^{-2}$}
    }
\startdata
TW Hya &  \phn7 &   337 & $\phantom{-}0.58 \pm 0.06$ & $ -0.12 \pm 0.08$ & $6.49 \pm 0.54$\\
\cdb   &  \phn8 &   325 & $ -0.22 \pm 0.09$ & $ -0.02 \pm 0.15$ & $4.72 \pm 0.48$\\
\hen   & 10 &   336 & $ -0.01 \pm 0.11$ & $ -0.06 \pm 0.16$ & $2.31 \pm 0.35$\\
\hd    & \phn7 &   330 & $  \phantom{-}0.06 \pm 0.23$ & $ -0.78 \pm 0.43$ & $5.69 \pm 1.33$\\
\cd    &  \phn8 &   122 & $ -0.31 \pm 0.10$ & $  \phantom{-}0.29 \pm 0.18$ & $4.40 \pm 0.76$\\
\enddata
\tablecomments{For comparison, hardness ratios for TTS in Taurus are HR 1 =
  0.3--1.0 (median = 0.89) and HR 2 = 0.0--1.0 (median = 0.25) for
  wTTS, HR 1 = 1.0 (for all but one cTTS) and HR 2 = 0.1--0.7 (median
  = 0.34) for cTTS.  (Neuh\"auser et al.\ 1995\markcite{n95}). }
\end{deluxetable}

We caution that the fluxes in Table \ref{table:rasspms} and in Kastner
\etal\ (1997\mc{k97}) for \hd, \cdb, and \hen\ are fluxes of unresolved
multiple systems.  There is no {\it a priori\/} way to know how to
divide the X-ray flux among the components.  Even a conclusion drawn
from the hardness ratios may be in error, as the individual
companions may have different ratios. 

The hardness ratios of the five stars in Table \ref{table:rasspms} can
be compared to typical values for TTS in Taurus.  Neuh\"auser \etal\ 
(1995\mc{n95}) found significantly different hardness ratios for
classical T Tauri stars (cTTS) compared to weak-line T Tauri stars
(wTTS), which they interpreted to be due to more absorption of soft
X-rays in cTTS due to their circumstellar material.

Adopting the usual dividing line of EW(H$\alpha$) $= 10$ \AA\ as
boundary between cTTS and wTTS (e.g., Bertout 1989\mc{b89}), TW Hya,
\hen, and \cd\ are cTTS while \hd\ and \cdb\ are wTTS.  However, as
noted by Kastner \etal\ (1997\mc{k97}), some authors consider the
presence of an infrared excess an additional criterion for
classification as a cTTS, and in the present case this criterion may
be more relevant as an influence on the X-ray emission.  Using only
the presence or absence of an infrared excess as the selection
criterion switches the classification of \hd\ from wTTS to cTTS and
\cd\ from cTTS to wTTS.

For cTTS in Taurus, Neuh\"auser \etal\ (1995\mc{n95}) found mean
hardness ratios of $\langle {\rm HR} 1\rangle = 0.994 \pm 0.027$ and
$\langle {\rm HR} 2 \rangle = 0.42 \pm 0.27$, where the errors given
are the standard deviations.  Hence, all five stars, regardless of
cTTS or wTTS status, show lower values, i.e.\ softer X-ray emission.
Indeed, several of the stars have negative values of HR1, softer than
any of the TTS observed in Taurus by \rne\ \etal\ (1995\mc{n95}).
This may be partly due to lower interstellar extinction, since the
stars are much closer than Taurus.  It also indicates the lack of
molecular gas in the immediate vicinity of these stars, either because
gas from their formation has dispersed or because they were formed
elsewhere.  TW Hya, however, is thought to be viewed pole-on (Kastner
\etal\ 1997\mc{k97}).  If so, even its substantial circumstellar disk
would not contribute strongly to X-ray absorption, yielding low
hardness ratios.  For the other stars, the low hardness ratios even in
those stars with infrared excesses may suggest that either they are
also viewed pole-on, or that their circumstellar material is less
attenuating than in Taurus TTS, possibly indicating disks with lower
masses and/or larger grains than those typically found around
cTTS\null.

Thus, the X-ray luminosity and variability of \hd\ and \cd\ are
consistent with those of PMS stars.  Their X-ray spectral properties
are somewhat different, perhaps reflecting the difference between
their current environment and that in Taurus-Auriga.

\section{Search for Additional Young Stars}\label{sec:othermembers}

It has been suggested that \hd\ and \cd, along with the other three
stars in Table \ref{table:rasspms}, are part of an isolated group
of young stars that formed together (Gregorio-Hetem \etal\ 
1992\mc{gh92}, Feigelson 1996\mc{f96}, Kastner \etal\ 1997\mc{k97}).
To test this idea, we searched for additional young stars in the area
using both our X-ray observations and observations from Hipparcos.

\subsection{Optical Counterparts to X-ray Sources}\label{sec:otherx}

Our X-ray detections of \hd\ and \cd\ were within 3\arcsec\ of their
optical positions, a typical value for the boresight offset of the HRI
(David \etal\ 1997\mc{d97}).  To search for optical counterparts to
the other X-ray sources, we used images of the HRI fields from the
Digitized Sky Survey (DSS; Morrison 1995\mc{m95}).  Most
previously-known X-ray sources observed by HRI were detected with
10\arcsec\ of their catalog positions (David \etal\ 1997\mc{d97}).
Therefore, we set 10\arcsec\ as an upper limit for the permitted
offset between the X-ray position and any candidate optical
counterparts.  The DSS images were examined visually to identify
optical sources within a 10\arcsec-radius circle centered on each
X-ray position.  For each X-ray source, only one optical counterpart
was found within this radius.  The coordinates of the optical sources
are given in Tables \ref{table:opt-hd} and \ref{table:opt-cd}.  All
optical counterparts appeared stellar (not extended) in the DSS images
except for that of RX J113220.7$-$343343 which is slightly elongated.

\begin{deluxetable}{ccccc}
\tablecaption{Optical counterparts in the HD 98800 Field \label{table:opt-hd}}
  \tablewidth{0pt}
  \tablenum{4}
  \tablehead{
    \colhead{X-ray source} & \colhead{R.A. (optical)}
    & \colhead{Dec. (optical)} & \colhead{Offset} & \colhead{$B_J$} \\
    \colhead{} & \colhead{(J2000)} & \colhead{(J2000)}
    & \colhead{(arcsec)} & \colhead{(mag)}
    }
\startdata
RX J112205.4$-$244641\tablenotemark{a} & $11^{\rm h}22^{\rm m}05\fs6$ & $-24\arcdeg46\arcmin39\farcs5$  & 2.6 & 10.1 $\pm$ 0.4\\ 
 RX J112303.2$-$243629   & $11^{\rm h}23^{\rm m}03\fs1$ & $-24\arcdeg36\arcmin27\farcs1$ &  2.9  & $> 18$ \\
 RX J112157.4$-$244014  & $11^{\rm h}21^{\rm m}57\fs1$ & $ -24\arcdeg40\arcmin15\farcs5$ &  4.8 & $> 18$ \\
 RX J112112.4$-$244032  & $11^{\rm h}21^{\rm m}12\fs6$ & $ -24\arcdeg40\arcmin34\farcs8$ &  2.9 & $> 18$ \\
 RX J112229.2$-$244258  & $11^{\rm h}22^{\rm m}29\fs1$ & $ -24\arcdeg42\arcmin59\farcs9$ &  2.8 & $> 18$ \\
 RX J112154.5$-$244413  &  $11^{\rm h}21^{\rm m}54\fs5$ & $ -24\arcdeg44\arcmin12\farcs8$ &  0.7 & $> 18$ \\
 RX J112204.4$-$245347  & $11^{\rm h}22^{\rm m}04\fs4$ & $ -24\arcdeg53\arcmin45\farcs1$ &  1.9 & $> 18$ \\
 RX J112156.1$-$245503  & $11^{\rm h}21^{\rm m}56\fs3$ & $ -24\arcdeg55\arcmin01\farcs5$ &  2.1 & $> 18$ \\
 RX J112216.6$-$245810  & $11^{\rm h}22^{\rm m}16\fs9$ & $ -24\arcdeg58\arcmin08\farcs7$ &  4.1 & $> 18$ \\
\enddata
\tablenotetext{a}{HD 98800}
\end{deluxetable}

\begin{deluxetable}{ccccc}
  \tablecaption{Optical counterparts in the CD $-$33$\arcdeg$7795
    Field \label{table:opt-cd}}
  \tablewidth{0pt}
  \tablenum{5}
  \tablehead{
    \colhead{X-ray source} & \colhead{R.A. (optical)}
    & \colhead{Dec. (optical)} & \colhead{Offset} & \colhead{$B_J$} \\
    \colhead{} & \colhead{(J2000)} & \colhead{(J2000)}
    & \colhead{(arcsec)} & \colhead{(mag)}
    }
\startdata
RX J113155.3$-$343630\tablenotemark{a} & $11^{\rm h}31^{\rm m}55\fs5$ & $ -34\arcdeg36\arcmin28\farcs1$   & 2.9 & 13.3 $\pm$ 0.6\\
 RX J113220.7$-$343343 & $11^{\rm h}32^{\rm m}21\fs1$ & $ -34\arcdeg33\arcmin49\farcs7$   & 7.7 & 17.1 $\pm$ 0.8\\ 
 RX J113126.2$-$343411 & $11^{\rm h}31^{\rm m}26\fs4$ & $ -34\arcdeg34\arcmin10\farcs1$   & 3.0 & $> 18$ \\ 
 RX J113143.3$-$343704 & $11^{\rm h}31^{\rm m}43\fs2$ & $ -34\arcdeg37\arcmin04\farcs0$   & 2.0 & $> 18$ \\ 
 RX J113213.6$-$344127 & $11^{\rm h}32^{\rm m}13\fs8$ & $ -34\arcdeg41\arcmin25\farcs6$  & 2.2 & $> 18$ \\ 
 RX J113140.5$-$344246 & $11^{\rm h}31^{\rm m}40\fs8$ & $ -34\arcdeg42\arcmin43\farcs6$  & 4.5 & $> 18$ \\ 
\enddata
\tablenotetext{a}{CD $-$33$\arcdeg$7795}
\end{deluxetable}

To measure magnitudes for the optical counterparts, we used the IRAF
{\it apphot} package.  The sky background level was measured at eight
places in each image and found to be constant to better than 1\%
across the image; the average of these eight sky values was then
subtracted from each image. An additional 500 counts were subtracted
from each image since the DSS photometric calibration is defined in
terms of the total signal that is more than 500 counts above the
background.  We then set all negative pixels in each image to zero and
measured the integrated signal for each source.  This signal was
converted to a SERC-$J$ magnitude using the photometric calibration
from the DSS.\footnote{The photometric calibration is available at
  http://www-gsss.stsci.edu/dss/photometry/index.html.  The southern
  sky images in the DSS are digitized from SERC-$J$ photographic
  plates (IIIa-J emulsion + GG395 filter).  The photographic $J$
  bandpass (not to be confused with the $\lambda=1.25$ \micron\ 
  Johnson $J$ band) is a blue bandpass similar to the Johnson $B$
  (see, e.g., Kron 1980\mc{k80}, Lasker \etal\ 1990 \mc{l90}).}  The
magnitudes are given in Tables \ref{table:opt-hd} and
\ref{table:opt-cd}.  All but one of the optical counterparts (besides
the central stars) were fainter than the $J = 18$ limit of the DSS
photometric calibration.  As a check, the $J$ magnitudes of \hd\ and
\cd\ were measured; these magnitudes agree with their {\em Guide Star
  Catalog} magnitudes (which come from the same plates) within the
errors.

The extreme faintness of the other optical counterparts strongly
suggests that few if any of them are pre--main-sequence stars
associated with \hd\ and \cd.  We compared their measured magnitudes
to those predicted by the evolutionary tracks of D'Antona \&
Mazzitelli (1994\mc{dm94}).  We used a SERC-$J$ to Johnson $B$
conversion of $B = J + 0.28(B - V)$ (Lasker \etal\ 1990\mc{l90}) and
used values of $(B - V)$ and bolometric corrections from Kenyon \&
Hartmann (1995\mc{kh95}) to convert the theoretical tracks to observed
magnitudes.  At a distance of 50 pc, a star with $J > 18$ would lie
below the main sequence if it has a spectral type earlier than
M2.5\null.  In order to be coeval with \hd\ ($10^7$ yr) and lie at the
same distance, a star with $J > 18$ would have to have a spectral type
later than M5\null.  While this could be true for individual stars in
our sample, it is extremely unlikely that any cluster of young stars
around \hd\ or \cd\ would contain {\em only} very-late-type stars
aside from the central stars.  The low X-ray fluxes of the other
sources (1--2 orders of magnitude fainter than \hd\ or \cd) further
indicates that it is unlikely that any of the other stars in these
fields are PMS stars.

Hoff \etal\ (1996\mc{hph96}, 1998\mc{hhp98}) used a similar technique
to search for additional young stars in the vicinity of TW Hya and
\cdb, with the same result.  Their deep \rosat\ PSPC observations and
follow-up optical spectroscopy revealed no additional young stars in
the $2\arcdeg$-diameter PSPC fields around each star.

While our deep HRI observations did not reveal any additional
pre--main-sequence stars, this does not rule out the presence of
additional stars in a cluster, since the area covered was fairly
small.  At 50 pc, each 40\arcmin\ HRI field covers an area projected
on the sky with a radius of roughly 0.3 pc.  The median projected
separation of young stars in the Taurus-Auriga star-forming region
(excluding close binary companions) is $\sim 0.3$ pc (Gomez \etal\ 
1993\mc{g93}).  Thus, it is perhaps not surprising that we did not
detect additional young stars in our HRI fields.  The PSPC
observations of Hoff \etal\ (1996\mc{hph96}, 1998\mc{hhp98}) cover
roughly three times the area of our HRI observations, suggesting that
any group in the TW Hya area may be less dense than Taurus-Auriga.
Clearly we need to probe a larger area in order to test for the
presence of a cluster of young stars.  In the following section, we
discuss our search for additional members of a putative group over a
larger area using Hipparcos data.

\subsection{Hipparcos Sources}\label{sec:hipparcos}

The biggest obstacle to identifying isolated young stars is typically
the uncertain distances to any candidates.  Without an associated
molecular cloud or cluster, the distance to a star is very uncertain
(unless it is close enough to have a measured parallax) and it cannot
be placed on an HR diagram to determine its age from theoretical
evolutionary tracks.  Instead, more indirect arguments must be used.
The identification of a distributed population of young stars via
X-ray emission and Li absorption has recently been called into
question by Brice\~no \etal\ (1997\mc{b97}), who point out the
similarities in X-ray and optical properties of stars with ages from
$10^6$ to $10^8$ yr.  The Hipparcos catalog (ESA 1997\mc{esa97}; see also
Perryman \etal\ 
1997\mc{p97}) alleviates this problem somewhat, since it contains
distances to over $10^5$ stars in the solar neighborhood, allowing
accurate age determinations of many more stars than was previously
possible.

We used the Hipparcos catalog to test the question of whether there is
an additional population of $\sim 10^7$ yr old stars in the vicinity
of the five previously known young stars in the TW Hya area identified
by Gregorio-Hetem \etal\ (1992\mc{gh92}).  We placed all of the stars
observed by Hipparcos in this region on an HR diagram, and identified
candidate \pms\ stars by choosing stars above the main sequence.
Because stars selected in this way can be either pre--main-sequence or
post--main-sequence or unresolved multiple ZAMS or MS stars, we also
studied the kinematics and X-ray properties of the candidate young
stars.

For a specific comparison, we took the properties of the Taurus-Auriga
molecular cloud complex as typical of low-density star forming
regions and used these properties to define a search region in the
Hydra area.  The known \pms\ stars in Taurus-Auriga are distributed
over an area roughly $14\arcdeg \times 14\arcdeg$ on the sky.  Within
this area, there are several distinct subgroups with radii of $\sim$
0.5--1 pc and internal velocity dispersions of $\le 1$--2 km s$^{-1}$
(Jones \& Herbig 1979\mc{jh79}, Gomez \etal\ 1993\mc{g93}).  The group
sizes are similar to the Jeans length for the associated dark clouds
and to the measured size of the clumps in the molecular gas,
suggesting group formation from gravitational instabilities (Gomez
\etal\ 1993\mc{g93}).

One of these small groups of stars placed at the distance ($\sim$ 50
pc) and age ($\sim$ 10$^7$ yr) of HD 98800 would look quite different
from its present day appearance in Taurus.  A velocity of 1 km
s$^{-1}$ corresponds to a distance traveled of 10.2 pc in 10$^7$ yr.
Thus, if it is not gravitationally bound, a group of this age at a
distance of 50 pc should have an angular radius of order 10\arcdeg.
The intrinsic proper motions of any group members should also be
relatively small; a lateral motion of 1 km s$^{-1}$ at 50 pc gives 4.2
mas yr$^{-1}$.  Indeed, the space motions of TW Hya and \hd\ in the
plane of the sky are only a few km s\per\ (see Table \ref{table:cands}
and \sec\ \ref{section:kinematics}).

We used 11$^{\rm h}15^{\rm m}$ $-$32$\arcdeg 20\arcmin$ (J2000), the
mean RA and Dec.\ of the five previously-known young stars in this
area, as the center of our search area; four of the five lie within
$\sim$ 5\arcdeg\ of this point, with HD 98800 at $\sim$ 8\arcdeg.  We
searched an area 10\arcdeg\ in radius and selected all stars from the
Hipparcos catalog with parallaxes in the range 15.5--24.3 mas.  This
is the range of the parallaxes of TW Hya and HD 98800 plus or minus
their 1$\sigma$ uncertainties, corresponding to distances of
41.2--64.5 pc.  Thus, our search encompasses a volume of space roughly
10 pc in radius, centered on the five previously known young stars in
this area.  Additional selection criteria (which in practice excluded
only a few stars) were the presence of a $B-V$ measurement in the
Hipparcos catalog and a parallax uncertainty of less than 8.1 mas.

\begin{figure}
\plotone{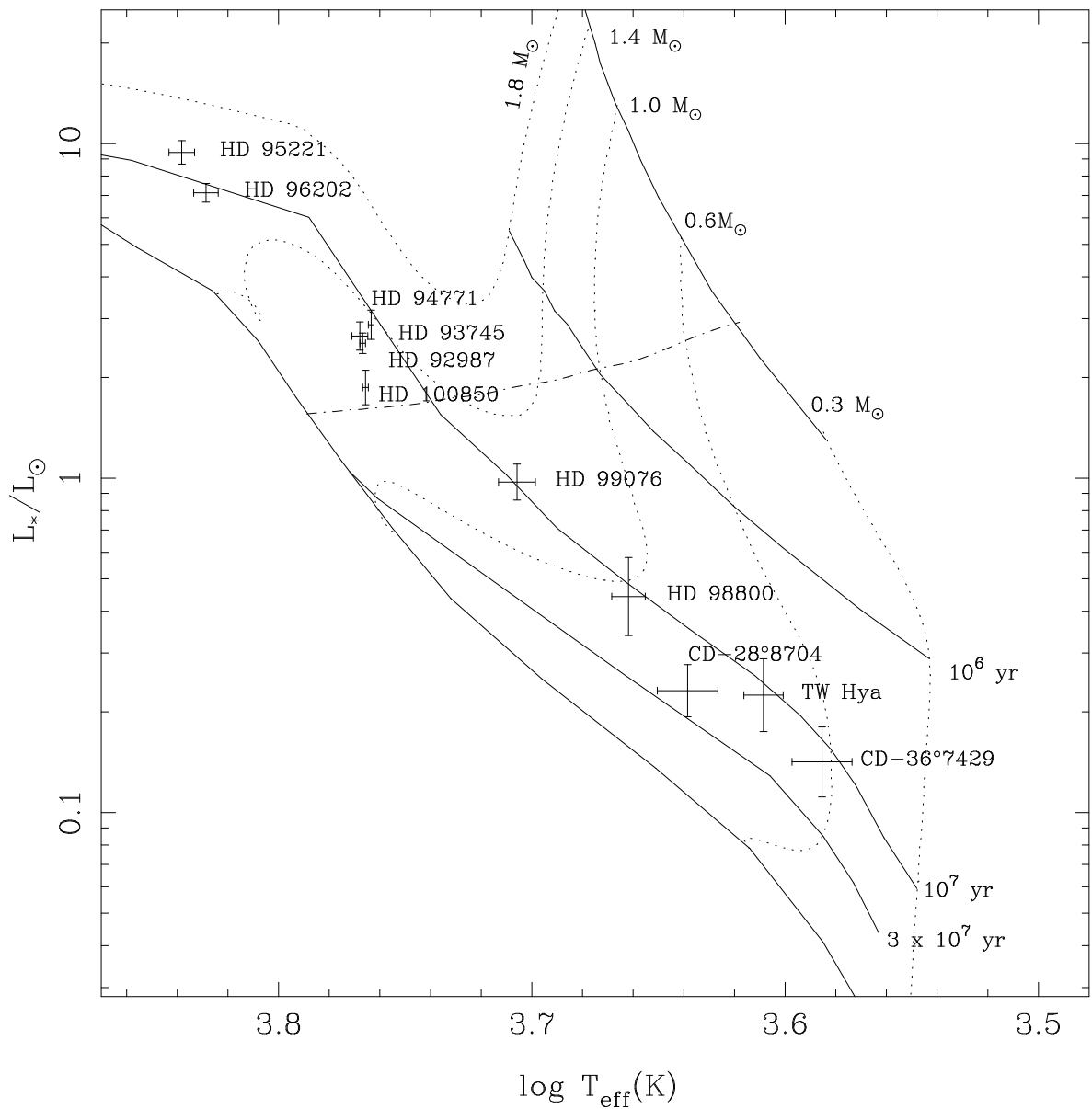}
\caption{HR diagram of candidate young stars.  The
  evolutionary tracks and isochrones are from D'Antona \& Mazzitelli
  (1994\protect\mc{dm94}) with Alexander opacities and CM convection.
  The dashed line shows $V = 7.8$ at a distance of 50 pc, with a
  bolometric correction added; the Hipparcos catalog is incomplete
  below this level.
  \label{figure:HR}
  }
\end{figure}

This search yielded 60 stars.  We rejected those with luminosity
classes I, II, or III (3 stars) and plotted the rest on an HR diagram.
Bolometric corrections and the conversion from spectral type to
effective temperature were taken from Kenyon \& Hartmann
(1995\mc{kh95}).  For a few stars, no spectral type was available, and
$B-V$ color was used instead.  We did not correct for extinction;
since the effective temperature for most of the stars was determined
from their spectral types rather than $B-V$ colors, the primary effect
of any extinction will be to make stars appear fainter than they
really are.  Since an unresolved binary companion can raise the
apparent luminosity of a star by up to a factor of two, we selected as
candidate young stars only those stars that lie more than a factor of
two above the zero-age main sequence (ZAMS).\footnote{In practice our
  criteria exclude most $10^7$-yr-old stars earlier than about F2,
  since a line a factor of two above the ZAMS intersects the 10$^7$ yr
  isochrone at $T_{\rm eff} \approx 6800$ K.}  This yielded 11 stars,
including HD 98800 and TW Hya, the only two of the five previously
known young stars that are in the Hipparcos catalog.  The candidate
young stars are listed in Table \ref{table:cands} and plotted in
Figure \ref{figure:HR}; we discuss them in detail below.

\begin{deluxetable}{llccc}
  \tablenum{6}
  \tablewidth{0pt}
  \tablecaption{Candidate young stars near TW Hya from Hipparcos
    \label{table:cands}}
  \tablehead{
    \colhead{Star} & \colhead{Spectral} & \colhead{Distance} &
    \colhead{Multiplicity}  & $V_{LSR}\tablenotemark{a}$ \\
    \colhead{} &\colhead{type} &\colhead{(pc)} &\colhead{} &
    \colhead{ (km s$^{-1}$)} 
    }
\startdata
  HD 95221 &   F2 V & 48.6 $\pm$ 2.0       & binary & 15.8\\
  HD 96202 & F3 IV/V & 43.5 $\pm$ 1.4      & binary & 25.4\\
  HD 93745\tablenotemark{b} &   G2 V & 55.5 $\pm$ 2.7       & single & 64.4\\
  HD 92987\tablenotemark{b} & G2/G3 V & 44.0 $\pm$ 1.5      & single & 16.2\\
 HD 100850\tablenotemark{b} &   G3 V & 59.9 $\pm$ 3.6       & single & 39.4\\
  HD 94771 & G3/G5 V & 54.7 $\pm$ 2.8      & single & \phn9.4\\
  HD 99076\tablenotemark{c} &   K1 V & 49.9 $\pm$ 3.1       & single & 33.5\\
  HD 98800 &   K4 V & 46.7 $\pm$ 6.2       & quadruple & \phn4.5\\
CD $-$28$\arcdeg$8704 &   K5 V & 50.6 $\pm$ 4.5 & single & 10.5\\
  TW Hya &   K7 V & 56.4 $\pm$ 7.0       & single & \phn3.9\\
CD $-$36$\arcdeg$7429 & Mp\tablenotemark{d} & 50.3 $\pm$ 6.0    & binary & \phn2.2\\
\enddata
\tablenotetext{a}{Transverse velocity relative to the local standard
  of rest based on Hipparcos proper
  motion and distance.}
\tablenotetext{b}{Classified as ``inactive'' by Henry et al.\
  (1996\markcite{h96}) based on \ion{Ca}{2} H and K emission.}
\tablenotetext{c}{Classified as ``active'' by Henry et al.\
  (1996\markcite{h96}) based on \ion{Ca}{2} H and K emission.}
\tablenotetext{d}{Spectral type of M0 used to plot on HR diagram.}
\end{deluxetable}

Four of the 11 stars are known multiples.  For each of these, we
determined the magnitude of the primary and used that to plot the
system on the HR diagram.  For CD $-$36$\arcdeg$7429 and HD 96202 we
used the primary magnitude from the Catalog of Components of Double
and Multiple Stars (CCDM; Dommanget \& Nys 1994\mc{ccdm}).  For HD
95221, we used the flux ratio and combined system magnitude from the
Hipparcos catalog to determine the individual component magnitudes.
For HD 98800, we used the primary (component Aa) magnitude of $V =
9.40$ from Soderblom \etal\ (1998\mc{s98}).  Correcting the magnitude
of HD 96202 places it less than a factor of two above the ZAMS; we
retain it in our sample for consistency in later comparisons with
other samples selected in the same way (see below), since it passed
the original selection criteria.

All but one of the stars have MK spectral types; for CD
$-$36$\arcdeg$7429, the CCDM gives a spectral type of Mp.  Since the
$B-V$ value ($1.66 \pm 0.4$) has such a large uncertainty, we adopted
a spectral type of M0 to place it on the HR diagram.  If the spectral
type is later than this, the inferred age will be even younger.

As Figure \ref{figure:HR} shows, there are several stars whose
positions on the HR diagram (if they are PMS stars) imply ages of 1--2
$\times 10^7$ yr, consistent with the ages previously found from
Hipparcos data for TW Hya (Wichmann \etal\ 1998\mc{w98}) and HD 98800
(Soderblom \etal\ 1998\mc{s98}). To further explore these stars'
properties, we examined the \rosat\ All-Sky Survey to determine their
X-ray fluxes.

Of the stars listed in Table \ref{table:cands}, two are clearly
detected in the RASS, namely CD $-$28$\arcdeg$8704 and CD
$-$36$\arcdeg$7429, while HD 99076 is marginally detected (at $\sim 2
\sigma$ significance).  The X-ray source found near CD
$-$28$\arcdeg$8704 might, however, be unrelated. The X-ray source is
1\arcmin\ off the optical position while most X-ray sources found near
the Taurus TTS are within 40\arcsec\ (\rne\ \etal\ 1995\mc{n95}), and
there are two other stars in the DSS image of the field that lie with
40\arcsec\ of the X-ray position.

We list the X-ray properties of the three X-ray detected stars in
Table \ref{table:otherx}, including the offset between the X-ray and
optical positions, the likelihood of existence ${\cal L}$ in the RASS,
the individual exposure times, and the broad band (0.1 to 2.4 keV)
count rates (background subtracted and corrected for vignetting). As
in the case for the other young stars in this area with RASS fluxes,
some spectral information can be obtained using the hardness ratios.
These hardness ratios are given in the following two columns.  Count
rates from RASS observations were converted to fluxes using the
hardness ratios as above in \sec\ \ref{sec:spectral}.  The final
column gives the X-ray luminosity.

CD $-$36$\arcdeg$7429 has hardness ratios and \Lx\ in the range of
values given for the young stars in Table \ref{table:rasspms}.  As
discussed in the following section, its transverse velocity is also
the closest to that of HD 98800 and TW Hya, suggesting a possible
physical relationship.  CD $-$28$\arcdeg$8704 and HD 96771 are fainter
in X-rays than the previously-known young stars near TW Hya.  However,
the X-ray emission levels of these stars are not unheard of for T
Tauri stars; Feigelson \etal\ (1993\mc{f93}) found a mean X-ray
luminosity of $\log L_X = 29.2$ for T Tauri stars in Chamaeleon.  The
fact that CD $-$28$\arcdeg$8704 and HD 96771 are detected only at the
2--3 $\sigma$ level indicates that their X-ray fluxes are near the
sensitivity limit imposed by the short exposure times of the RASS,
which could account for the non-detection of some of the other stars.
However, the fact that none of the G stars in Table \ref{table:cands}
were detected in X-rays suggests that they may be older, since PMS G
stars tend to have higher X-ray luminosities than PMS stars of later
spectral types (Feigelson \etal\ 1993\mc{f93}, Gagn\'e \etal\ 
1995\mc{gcs95}, \rne\ \etal\ 1995\mc{n95}).

\begin{deluxetable}{lcrccccccc}
  \scriptsize
  \tablewidth{0pt}
  \tablenum{7}
  \tablecaption{X-ray properties of candidate young stars near TW Hya
    \label{table:otherx}
    }
  \tablehead{
    \colhead{Name} & \colhead{Offset} &
    \colhead{$\cal L$} & \colhead{Exposure}  & \colhead{counts ks$^{-1}$} &
    \colhead{HR 1} & \colhead{HR 2} & \colhead{Flux/$10^{-12}$ } &
    \colhead{$\log L_{X}$}\\
    \colhead{}  & \colhead{} & \colhead{}     &
    \colhead{Time (s)} & \colhead{}  & \colhead{} &
    \colhead{}
    & \colhead{erg s$^{-1}$ cm$^{-2}$} & \colhead{(cgs)} }
\startdata
CD $-$28$\arcdeg$8704 & 63\arcsec & 13 & 322 & $\phn39 \pm 14$ &
$-0.03 \pm 0.35$ & $0.03 \pm 0.49$ & $0.32 \pm 0.11$ & 29.0\\
HD 99076          & 30\arcsec & 5 & \phn95 & $\phn43 \pm 28$ &
$\phantom{-}0.77 \pm 0.66$ & $\ge -0.21$ & $0.53 \pm 0.42$ & 29.2\\
CD $-$36$\arcdeg$7429 & 19\arcsec & 115 & 133 & $345 \pm 54$ &
$-0.19 \pm 0.16$ & $-0.23 \pm 0.25$ & $2.52 \pm 0.61$ & 29.9\\
\enddata
\end{deluxetable}

The stars as a group do not show evidence of enhanced chromospheric
activity in optical observations.  Five of them were observed in the
\ion{Ca}{2} H and K lines by Henry \etal\ (1996\mc{h96}).  Of these,
three were classified as ``inactive''; only HD 99076 was classified as
``active''.  HD 98800 was observed but was not classified in either of
these groups due to the difficulty of determining the underlying
photospheric level near the H and K lines for cooler stars; it has
\ion{Ca}{2} H and K emission that is stronger than most field stars of
its color (Soderblom \etal\ 1996\mc{s96}).

The lack of strong X-ray emission from most of these stars suggests
that some of them may be $10^9$-yr-old subgiants that are just leaving
the main sequence rather than $10^7$-yr-old PMS stars.  However, early
F stars have very weak X-ray emission even in the PMS phase, so this
alone does not reveal whether these stars are young or old.  To
further investigate the origin of these stars we examined their
kinematics and compared this field with other fields at the same
Galactic latitude.  The kinematics not only help discriminate between
old and young stars; they also allow us to test the suggestion that
the young stars near \hd\ formed together as a group, perhaps from a
rapidly-moving cloudlet (Feigelson 1996\mc{f96}, Kastner \etal\ 
1997\mc{k97}).

\subsubsection{Kinematics of the young stars}\label{section:kinematics}

\begin{figure}
  \plotone{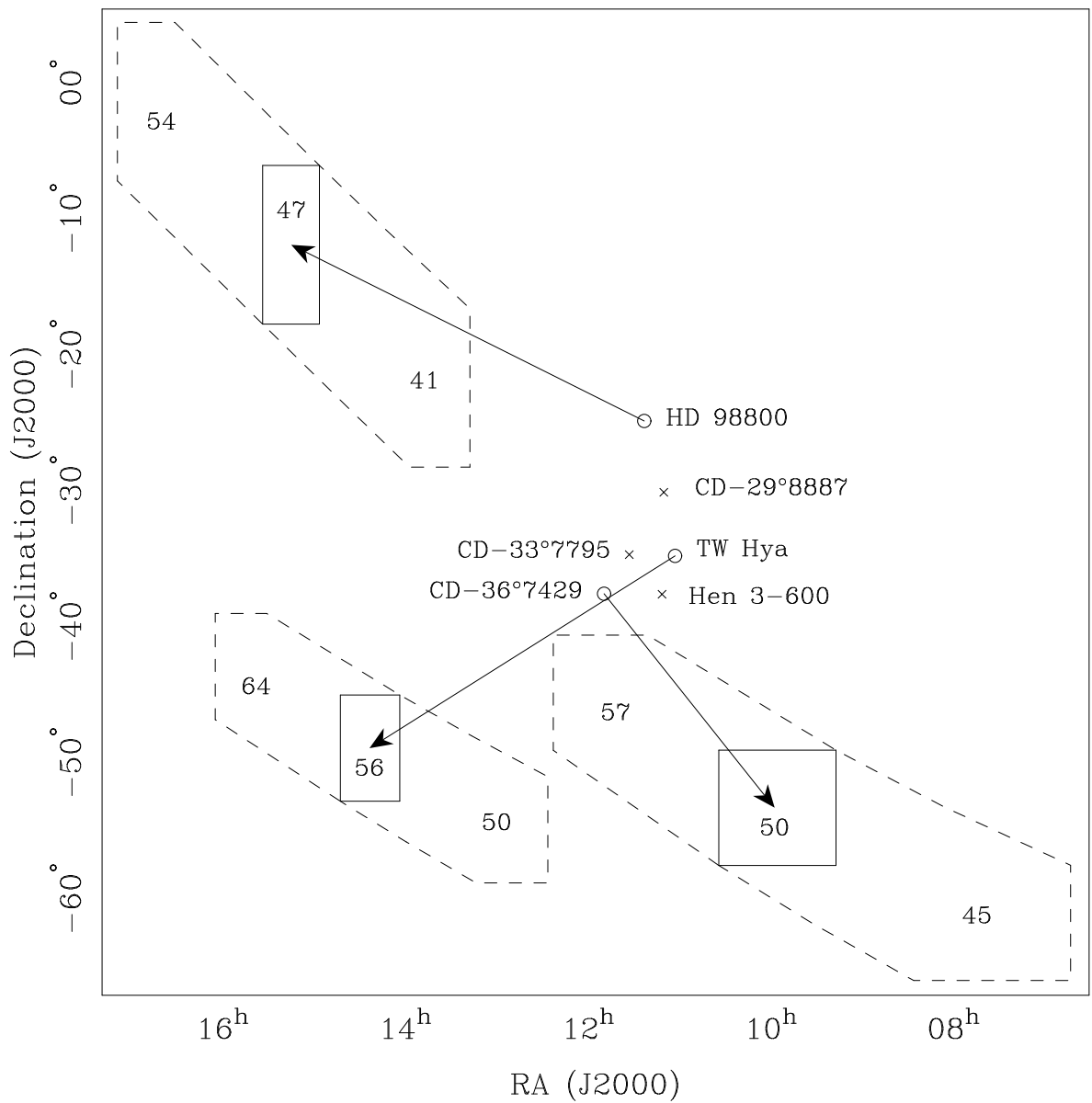}
\caption{
  Proper motions of confirmed young stars, projected back in time from
  the stars' current positions.  The tips of the arrows show the
  stars' positions on the sky $10^7$ yr ago; the boxes show the
  uncertainty on the proper motion.  If the stars formed together, the
  arrows should point toward their common place of origin.  This
  appears not to be the case.  Note that the length of the arrows
  represents motion over the estimated $10^7$ yr ages of TW Hya and HD
  98800.  Young stars without known proper motions are shown with
  crosses.  The dashed lines show the effect of $1 \sigma$ distance
  uncertainties on the subtracted solar motion.  The numbers represent
  the current distances in pc used to calculate the motions.
  \label{figure:pm}
}
\end{figure}

If the young stars near TW Hya were formed in a group from the same
molecular cloud, they should have similar space motions.  To test
this, we subtracted the motion of the Sun relative to the Local
Standard of Rest (LSR) from the observed proper motion of each
star.\footnote{We used an LSR velocity of $(U,V,W) = (9,12,7)$ km
  s\per\ (Delhaye 1965\mc{d65}), where $U$, $V$, and $W$ are positive
  in the directions of the Galactic center, Galactic rotation, and
  North Galactic Pole, respectively.}
The resulting motions are shown in Figure \ref{figure:pm}; the proper
motions of the stars have been projected backward in time to show the
positions of the stars on the sky relative to each other $10^7$ yr
ago. It is clear that over this amount of time, roughly their
estimated ages, the stars were not near each other.  Thus it
seems unlikely that they formed together.

All of the stars besides \hd, TW Hya, and CD $-$36$\arcdeg$7429 have
transverse space velocities greater than 10 km s\per. Even ignoring
the radial component of their motions, their relative velocities
(Table \ref{table:cands} and Figure \ref{figure:pm}) are much greater
than the few km s\per\ velocity dispersion seen for young stars in
Taurus (e.g., Jones \& Herbig 1979\mc{jh79}).  These high velocities
are more characteristic of giants than of young stars.  In contrast,
CD $-$36$\arcdeg$7429 has a space velocity similar to that of \hd\ and
TW Hya, and it is the brightest X-ray source of the group.  Its X-ray
brightness and low space velocity suggest that it may well be a \pms\ 
star.

We conclude that most of the stars in Table \ref{table:cands} are
probably not young (being perhaps mis-classified subgiants or giants,
or unresolved triple systems), while CD $-$36$\arcdeg$7429 is quite
likely a $10^7$-yr-old PMS star.  As this paper was nearing
completion, we obtained spectra of these stars as part of another
observing program.  These observations confirm the conclusions above:
of the candidate young stars in Table \ref{table:cands}, only CD
$-$36$\arcdeg$7429 has detectable Li absorption; the others all have
EW(Li) $< 0.1$ \AA.  We will report in detail on these spectroscopic
follow-up observations in a later paper, where we shall also present
the radial velocities.

\subsubsection{Comparison with other fields}\label{section:otherfields}

Is there an excess of young stars near HD 98800?  We searched other
fields at similar Galactic latitudes to find out.  The center of our
original search field corresponds to Galactic coordinates $(l,b) =
(280\fdg3,+26\fdg3)$.  We selected 28 other fields centered at $|b| =
26\fdg3$, starting at $l = 16\arcdeg$ and separated by increments of
$l = 24\arcdeg$ (which gives an angular separation of 21\fdg4).
Thus the edges of these 10\arcdeg-radius fields are separated by
roughly 1\fdg4 and cover most of the sky at $|b| = 26\fdg3$.  The only
fields excluded were (280\arcdeg, 26\fdg3), which overlaps our
original field, and (184\arcdeg,$-$26\fdg3), which includes the
Hyades.

We applied the same selection criteria used for the original field
(parallax of 15.5--24.3 mas, parallax uncertainty $< 8.1$ mas, and
presence of a $B-V$ measurement) to each of these fields.  As before,
we rejected stars with luminosity classes I, II and III, plotted the
remaining stars on the HR diagram, and counted the number of stars in
each field that were more than a factor of two above the ZAMS.

First, we simply compared the number of stars meeting our selection
criteria without knowledge of whether these stars are pre-- or
post--main-sequence.  The average for 28 fields is 11.4 stars, with a
standard deviation of 3.9, not significantly different than the 11
stars found in our main search field.  Limiting the comparison to only
stars later than G0 (to match the range of spectral types for T Tauri
stars) and eliminating any stars with spectral class IV gives 7.2
$\pm$ 3.5 stars per field in the comparison fields, compared with 9 in
our main search field, again not a significant difference; similar
results are found for stars K0 and later.  Even adding \cd, \cdb, and
\hen\ to our main search field sample and comparing against the
Hipparcos-only sample in other fields gives only a 1$\sigma$ excess of
stars K0 and later. Comparing the percentages of stars meeting these
criteria rather than the absolute numbers does not change the results.

Thus, the presence of 11 Hipparcos-selected stars above the main
sequence in the field near \hd\ represents not an unusual excess but
rather the norm in a 10\arcdeg-radius field for that Galactic latitude
and distance.  However, if most of the stars selected in this way are
post--main-sequence stars rather than pre--main-sequence stars, as
suggested by the X-ray emission and kinematics of the candidates in
our main search field, then this comparison does not tell us whether
the young-star population near TW Hya is unique.

As noted above, \hd, TW Hya, and the X-ray bright CD
$-$36$\arcdeg$7429 have lower transverse space velocities relative to
the LSR than do most of the other stars in their field.  Thus, we also
compared the number of stars with low transverse space velocities in
our main field to that in other fields.  We counted the number of
stars in each field that meet our initial selection criteria and also
have $V_{LSR} \le 5$ km s\per.  In our main search field, there are
three.  Of the other 28 fields, 22 have no candidates with velocities
this low, five have one candidate, and one has two candidates.  This
gives an average of 0.35 per field (including our main field).  Given
this average rate, the Poisson probability of three or more low
velocity stars in a field is 0.005.  Thus, the main search field does
have a low-velocity population that is significantly above the mean.
We note that we had no {\it a priori\/} knowledge of the stars' space
velocities in choosing this field.  However, we caution that we did
not choose the velocity cutoff {\it a priori\/} though it is
physically plausible based on velocity dispersions of known young
stars.  If low transverse velocity is a reliable discriminant between
old and young stars, as the close relationship between space velocity,
X-ray flux, and Li shown above suggests, the area around TW Hya does
indeed seem to have significantly more young stars than other fields
at the same distance and Galactic latitude.

\section{Discussion}\label{sec:discussion}

Feigelson (1996\mc{f96}) suggested that the five young stars in the
vicinity of \hd\ and TW Hya might have formed from a rapidly-moving
cloudlet that has since dissipated.  In this model, the young stars
might have high space velocities but should have a relatively low
velocity dispersion among them.  This hypothesis is ruled out for \hd\ 
and TW Hya by the very different direction of their space motions.

Two of the young stars, TW Hya and CD $-$36$\arcdeg$7429, apparently
were relatively close to each other (at least in projection) a few
million years ago.  Without knowing their radial velocities it is
difficult to tell how close they were.  However, extrapolating their
present motions to their birth epoch of roughly $10^7$ years ago
places them far apart.

It is still possible that one or more of the other three stars (\cd,
\cdb, and \hen) shares a common origin with \hd, CD
$-$36$\arcdeg$7429, or TW Hya.  In addition, it is likely that more
young, late-type stars exist in this area.  The completeness limit of
the Hipparcos catalog depends on Galactic latitude and spectral type;
it is $V = 7.8$ for stars later than G5 and $V = 8.4$ for stars
earlier than or equal to G5 at $b = 26\fdg3$ (Turon \etal\ 
1992\mc{t92}).  Even taking the faint end of this completeness limit
(the dashed line in Figure \ref{figure:HR}), the Hipparcos catalog is
seriously incomplete for K and M stars at a distance of 50 pc.
However, until the proper motions of these fainter stars can be
measured, there is no evidence that a group of \pms\ stars formed
together in this area.

This fact, combined with the statistical analysis in the previous
section, leads to a puzzling result.  There seems to be an excess of
young stars in this area, and yet their space velocities suggest that
at least some of them did not form together.  Studies of the
kinematics of the other young stars in this area may help resolve
this apparent paradox.

Finally, we note that the apparent excess of young stars in this area
from the work of de la Reza \etal\ (1989\mc{dlr89}) and Gregorio-Hetem
\etal\ (1992\mc{gh92}) was discovered partly by chance.  Neither \cdb\ 
nor \cd\ meet the original selection criteria (an {\it IRAS\/}
detection) of these studies.  After \cdb\ was observed
spectroscopically, the \iras\ source nearby that led to its original
selection was found to be associated with a galaxy projected nearby.
\cd\ was identified later because of its spectroscopic similarity to
\cdb.  While these two stars are most probably \pms\ based on their
X-ray properties, there may be many more such sources in areas that
have not been surveyed spectroscopically in as much detail.

\section{Summary and Conclusions}\label{sec:summary}

We have presented pointed \rosat\ observations of the isolated young
stars \hd\ and \cd.  Their X-ray fluxes and X-ray variability are
consistent with them being \pms\ stars.  No other \pms\ stars were
found among X-ray sources within 40\arcmin\ of these stars.

Hipparcos observations of the area reveal the presence of nine
additional stars that lie above the main sequence in a volume roughly
10 pc in diameter centered on the five previously known young stars.
The X-ray properties and kinematics of these stars indicate that one
of them (CD $-$36$\arcdeg$7429) is quite likely a PMS star with
an age of $10^7$ yr, while the others are more likely
post--main-sequence stars. Observations of Li abundances in these
stars confirm this conclusion.

Comparison with other fields at the same Galactic latitude shows that
other fields selected in the same way show similar numbers of stars
above the main sequence.  However, the field around TW Hya has
significantly more stars with low space velocities (less than 5 km
s\per\ relative to the LSR) than the other fields.  Thus, there is
some indication of an excess of young stars in this area. Nonetheless,
the proper motions of \hd, TW Hya, and CD $-$36$\arcdeg$7429, if
extrapolated back in time, do not indicate a common place of origin.

\noindent \centerline{\bf Note added in proof:}

Our analysis of whether HD 98800, CD $-$36$\arcdeg$7429, and TW Hya could
have formed as a group neglected an important source of uncertainty.
The uncertainty in distance to these stars, while it does not affect
the observed proper motions, does introduce uncertainty into the solar
motion that must be subtracted in order to determine the stars'
intrinsic proper motions with respect to the LSR.  The 1$\sigma$
effect of this uncertainty is shown by the dashed lines in Figure 3.
Including this uncertainty indicates that TW Hya and CD
$-$36$\arcdeg$7429 could plausibly have formed together; only a
$1.5\sigma$ error on their distances is required. Regarding HD 98800,
bringing its current distance $2\sigma$ closer, to 36.8 pc, would
allow its projected point of origin to overlap with those of the other
two stars.  However, this distance makes HD 98800 a main-sequence star, in
conflict with its observed properties and, more importantly, of a very
different age than the other two stars.  We conclude that HD 98800 is
unlikely to share a common origin with TW Hya or CD
$-$36$\arcdeg$7429.

\section*{Acknowledgments}

We thank the referee, David Soderblom, for insightful comments
that improved the paper.
Thanks to Bob Mathieu, Guillermo Torres, David Koerner, and Andrea
Schweitzer for useful discussions, and to David Soderblom, Rainer
Wichmann, and Willi Hoff for providing results in advance of
publication.  Eric Jensen gratefully acknowledges the support of NASA
grant NAG5-3315 and the National Science Foundation's Life in Extreme
Environments Program through grant AST-9714246.  Ralph Neuh\"auser
acknowledges grants from the Deutsche Forschungsgemeinschaft
(Schwerpunktprogramm `Physics of star formation').  The ROSAT project
is supported by the Max-Planck-Gesellschaft and Germany's federal
government (BMBF/DLR).  This work made use of the Simbad database,
operated at CDS, Strasbourg, France, and the Digitized Sky Surveys,
produced at the Space Telescope Science Institute under U.S.
Government grant NAG W-2166.  The images of these surveys are based on
photographic data obtained using the Oschin Schmidt Telescope on
Palomar Mountain and the UK Schmidt Telescope.  The plates were
processed into the present compressed digital form with the permission
of these institutions.  The UK Schmidt Telescope was operated by the
Royal Observatory Edinburgh, with funding from the UK Science and
Engineering Research Council (later the UK Particle Physics and
Astronomy Research Council), until 1988 June, and thereafter by the
Anglo-Australian Observatory. The blue plates of the southern Sky
Atlas and its Equatorial Extension (together known as the SERC-J), as
well as the Equatorial Red (ER), and the Second Epoch [red] Survey
(SES) were all taken with the UK Schmidt.

\end{document}